\documentclass[aps,pre,superscriptaddress,twocolumn,showpacs,floatfix]{revtex4}
\usepackage{epsfig}

\newcommand{\dist}{{\rm dist}}
\newcommand{\be}{\begin{equation}}
\newcommand{\ee}{\end{equation}}
\newcommand{\bc}{\begin{center}}
\newcommand{\ec}{\end{center}}
\newcommand{\bi}{\begin{itemize}}
\newcommand{\ei}{\end{itemize}}
\newcommand{\ba}{\begin{eqnarray}}
\newcommand{\ea}{\end{eqnarray}}

\newcommand{\ignore}[1]{}

\begin{document}

\title{Effective dimensions and percolation in hierarchically
structured scale-free networks}

\author{V\'{\i}ctor M. Egu\'{\i}luz}
\affiliation{
Instituto Mediterr\'aneo de Estudios Avanzados IMEDEA (CSIC-UIB),
E07071 Palma de Mallorca, Spain}
\author{Emilio Hern\'andez-Garc\'{\i}a}
\affiliation{
Instituto Mediterr\'aneo de Estudios Avanzados IMEDEA (CSIC-UIB),
E07071 Palma de Mallorca, Spain}
\author{Oreste Piro}
\affiliation{
Instituto Mediterr\'aneo de Estudios Avanzados IMEDEA (CSIC-UIB),
E07071 Palma de Mallorca, Spain}
\author{Konstantin Klemm}
\affiliation{Niels Bohr Institute, CATS, Blegdamsvej 17, DK-2100 Copenhagen
\O, Denmark}

\date{\today}

\begin{abstract}

We introduce appropriate definitions of dimensions in order to
characterize the fractal properties of complex networks. We compute
these dimensions in a hierarchically structured network of particular
interest. In spite of the nontrivial character of this network that
displays scale-free connectivity among other features, it turns out to
be approximately one-dimensional. The dimensional characterization
is in agreement with the results on statistics of site percolation and
other dynamical processes implemented on such a network.

\end{abstract}

\pacs{89.75.Hc, 05.10.-a}

\maketitle

\section{Introduction.}
The most basic characteristic of any geometric structure is perhaps its
dimensionality. The notion of dimension is intuitively associated with
the amount of data necessary to locate a point on the structure, but
the difficulties to formalize this association are known since more
than a century ago \cite{Jackson}. A distinction should be made between
definitions of dimensions based on topology and those based on measures
and distances or metrics. While both types are important in different
fields of science, the latter is very relevant in the description of
fractal structures \cite{Mandelbrot} and in dynamical systems theory
\cite{Ott}.

Complex networks are a category of geometrical structures that has been
thoroughly investigated in the last few years\cite{Albert02}.
However, the possible characterization of complex networks in terms of suitably
defined dimensions remains practically unexplored, with the exception of a few
cases of networks constructed from or embedded in regular Euclidean lattices
\cite{Newman99,Rozenfeld02}. This characterization should not only improve
understanding of the different geometrical properties of various networks,
but also clarify its impact on the dynamics of processes that might take
place on them---percolation, disease propagation, information transmission,
etc. Fractal dimensions, for example, might be useful to elucidate the
connections between network topology correlations and dynamics which,
apart from some isolated results \cite{Boguna02}, remain essentially
not understood. Issues such as why two networks with the same degree
distribution but different wiring details show different dynamical
properties \cite{Eguiluz02} are good candidates to be tackled with the tools
of fractal geometry.

The aim of this Paper is to introduce a set of quantities, namely, the {\em
network dimensions} with the purpose of providing a finer
classification of networks with similar topological structure. As an
application, we analyze a particular type of structured scale-free network
\cite{Klemm02a,Klemm02b}. We show that some of its properties can be
understood from the fact that it behaves close
to one-dimensional with appropriately defined dimensions.

\section{Definitions.}
A network is a set of lines, the links, connecting points named nodes
or sites. Topologically, any network is a one-dimensional object that
by virtue of Whitney's theorem \cite{Jackson} can always be embedded in
the three-dimensional Euclidian space. However, this topological
dimension does not carry information related to the many interesting
properties of networks. In a regular square lattice, for example, the
number of sites within a given distance from a particular node
asymptotically grows as the square of this distance. This example
suggests that a metric definition of dimension could highlight better
the two-dimensional character of the lattice than the topological
dimension does.

A hierarchy of definitions for the dimension of a set that are
associated to the properties of
measures defined on that set were long
ago introduced by Renyi \cite{Renyi70}, and have since been used with
success in several
fields. These dimensions were particularly helpful for the description
of several natural fractal objects as well as for the characterization of
chaotic trajectories in
dynamical systems theory. In the standard definition the set to be
characterized is first covered
with a number $N(\epsilon)$ of boxes of size $\epsilon$.
Let $\mu_i$ be the measure associated to the box $i$. Then, the spectrum of
dimensions $\tilde D_q$ is defined by the scaling of the
quantity $\Gamma(q,\epsilon) \equiv \sum_{i=1}^{N(\epsilon)}
\mu_i^q$ for small $\epsilon$:
\be
\tilde D_q=\lim_{\epsilon\rightarrow 0} \frac{1}{q-1}{\ln
\Gamma(q,\epsilon) \over \ln \epsilon }  \ \ .
\label{renyi}
\ee
$\tilde D_0$, $\tilde D_1$, and $\tilde D_2$ are
the so-called capacity, information, and correlation dimensions respectively.
It can be shown that $\tilde D_q \geq \tilde D_{q'}$ if $q<q'$.
$\tilde D_1$ in Eq.~(\ref{renyi}) can in principle depend on the
particular set of boxes covering the set which implies that an extremum
requirement similar to the one in the original definition by
Haussdorf \cite{Ott} may be technically needed.
Here we will postpone the consideration of these
refinements to simplify implementation of
practical numerical algorithms.

\begin{figure}
\centerline{\epsfig{file=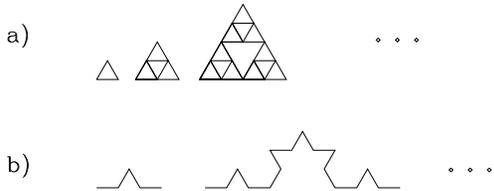,width=.45\textwidth}}
\caption{a) The three initial steps in the construction of an
infinite network of the Sierpi{\'n}sky gasket type. The unit of
length is kept to one link at each step of the construction, so
that the relevant scaling is at large distances. b) The two
initial steps in the construction of the nonbranching Koch-like
network.}
\label{fig:SK}
\end{figure}

One natural way to define a measure on a network is to assign
the unit of mass to
each node. In this case, the measure of a portion of the
network is the number of nodes that it contains. It is less easy to
define a `covering' of the network with boxes because it
requires an {\em a priori} knowledge of the Euclidean space in
which the network can be embedded. Since many networks are defined without
referencing any embedding in a Euclidean space,
alternative definitions based on intrinsic distances are necessary.
In our context, a convenient {\em distance} $\dist(X,Y)$ between two
nodes $X$ and $Y$ is the minimum number of links contained
by a path connecting nodes $X$ and $Y$. This is a
well-defined metric sometimes called {\em chemical distance}
\cite{Havlin84}. Distinctively from Euclidian and other
distances commonly used to define dimensions, this one is integer
valued. As a consequence, the scaling properties should be referred
to the large distances limit instead of the opposite one usually
considered.

A definition of the dimensions spectrum ${D_q}$,
equivalent to Renyi's one
(\ref{renyi}) in those systems where both are
applicable, is based on the scaling of the $q$-correlation
function $C(q,L)$ \cite{HenProc83}. In a network of $M$ points,
$C(q,L)$ is the number of $q$-tuplets of points in the
network with mutual distances smaller than $L$, and divided by
$M^q$. We then have:
\be
D_q=\lim_{L\rightarrow \infty} \frac{1}{q-1}{\ln C(q,L) \over \ln
L}
\label{dimcorr}
\ee
The advantage of using (\ref{dimcorr}) instead of (\ref{renyi})
with the Euclidean distance replaced by the chemical one is
that (\ref{dimcorr}) does not require any box covering and therefore
no {\em a priori}
knowledge of the Euclidean embedding is necessary.

\begin{figure}
\centerline{\epsfig{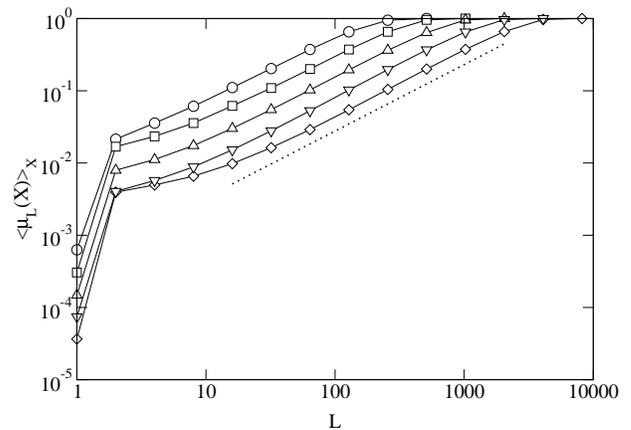}}
\caption{
Average number of nodes at distance $L$ from a given node. System sizes
10000, 20000, 40000, 80000, 160000. The dashed line grows as $\sim
L^{0.92}$. The averages have been done over 10 different networks and on
each 1000 different starting nodes.
} \label{fig:d2}
\end{figure}

Finally, another equivalent definition \cite{Benzi84} with easy
practical implementation is the one based on the scaling of the
number of neighbors within a given distance $L$ of a given site
$X_i$: $\mu_L(X_i) \equiv (M-1)^{-1}\sum_{j\neq i} \Theta( L -
\dist(X_i-X_j))$. $D_q$ is determined from moments of
$\mu_L(X_i)$:
\be
D_q=\lim_{L\rightarrow \infty} \frac{1}{q-1}{\ln \left<
\mu_L(X)^{q-1} \right>_X \over \ln L}  \ \ ,
\label{moments}
\ee
where the averages $\langle.\rangle_X$ are taken over all the
nodes $X$ in the network. Application of l'H{\^o}pital's rule
gives
\be
D_1=\lim_{L\rightarrow \infty} \left<{ \ln\mu_L(X)  \over \ln L}
\right>_X\ \ .
\label{D1}
\ee

Other equivalent definitions, based on the scaling of
nearest-neighbor distances, or fixed mass methods
\cite{vandeWater8895}, could be also implemented on networks, but
we find definition (\ref{moments}) to be appropriate for our
purposes. In cases in which an Euclidean distance can be defined,
it can also be used in (\ref{dimcorr}) or (\ref{moments}) to
define quantities that would be denoted by low-case letters,
$d_q$, to distinguish them from the case in which the chemical
distance is used. The $d_q$ are essentially the classical fractal
dimensions, but associated to the large-distance scaling.

It is instructive to calculate the dimension values for several
simple networks. For example, for regular triangular, square, etc.
lattices, it is easy to check that $D_q=2$ $\forall q$. $D_q=3$
for the classical three-dimensional lattices (e.g. cubic, ...),
etc. For a network with a {\em star} topology (i.e., a large
number $N$ of nodes connected to a central hub), $D_q = \infty$,
since all the nodes are at a finite distance (1 or 2) of each other
even in the limit $N\rightarrow \infty$. The same happens for
randomly wired networks. The usual implementation of the {\em
small world} property, i.e. the introduction of links connecting
arbitrarily remote nodes, leads also to $D_q \rightarrow \infty$.
This has been explicitly demonstrated for a small-world model in
\cite{Newman99}, where effective dimensions essentially equivalent
to $D_2$ are introduced and calculated as a function of spatial
scales. At large scales ($L\rightarrow\infty$ in our notation) a
divergent quantity is obtained. Thus, it is useful to think of
some of the characteristics of small worlds or random graphs as
being associated to an infinite dimensionality. From the
expressions in \cite{pseudofractal} or \cite{Jung2002} one can
also show that $D_q=\infty$ for the structures presented there.

\begin{figure}
\centerline{\epsfig{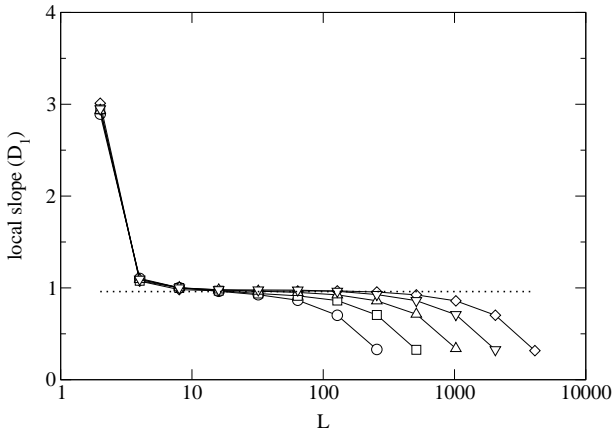}}
\caption{Average local ratio $\langle \ln(\mu_L)/\ln L\rangle$. The
plateau indicates a value $D_1 = 0.97$. Symbols and averaging as in
Fig.~\protect\ref{fig:d2}.
} \label{fig:d1}
\end{figure}

After considering these examples, one may wonder if there is any
non-regular complex network structure characterized by finite
dimensions. We will show below that the answer is positive and
that this finite dimensionality has dynamical consequences that
distinguish processes occurring on these networks from the
infinite-dimensional ones. Before that, we can mention that
networks of arbitrary dimensionality can be constructed by
following the rules used to construct recursively fractals of
given dimension. For example, a classical fractal is the
Sierpi{\'n}sky gasket \cite{Mandelbrot}, constructed, generation
after generation, by inscribing triangles inside the triangles
originated in the previous generation. Since we are here
characterizing large-scale features, it is better to consider the
construction as the recursive joining of triangles to construct a
larger and larger object (see Fig. \ref{fig:SK}a). Since the
resulting fractal structure is embedded in the plane, one can use
the Euclidean distance and find as usual that $d_q=d_0=\ln 3/\ln
2, \forall q$. If we use instead the chemical distance, with the
lines in Fig.~\ref{fig:SK}a identified as links of unit chemical
length, we have also $D_q=d_q=\ln 3/\ln 2, \forall q$. The
situation is rather different for {\em non-branching} fractal
constructions. For example, the classical Koch curve (Fig.
\ref{fig:SK}b) has $d_q=\ln 4/\ln 3$. But in terms of the chemical
metrics, $\dist(X,Y)$, the distance between two points is always
the number of nodes in between along the curve, so that we have
$D_q=1$ $\forall q$, as for any other non-branching structure. In
general, for networks constructed from the node and link structure
of a classical fractal object, one expects $D_q \leq d_q$ since
the Euclidian distance is in general shorter than the chemical
one. This generally leads to finite chemical dimensions. In other
constructions, the inequality may be reversed \cite{Rozenfeld02}.
This last reference also provides an additional example of a network
with finite dimensionality (in this case embedded in a Euclidean
lattice).

More interesting is the fact that one can construct networks with
finite dimensionality without any reference to fractal objects nor
to Euclidean embedding spaces. We show here that the networks
introduced by Klemm and Egu{\'\i}luz \cite{Klemm02a,Klemm02b} are
of such kind. This helps to understand the distinct behavior of
these structures as compared to other considered in the
literature.

\begin{figure}
\centerline{\epsfig{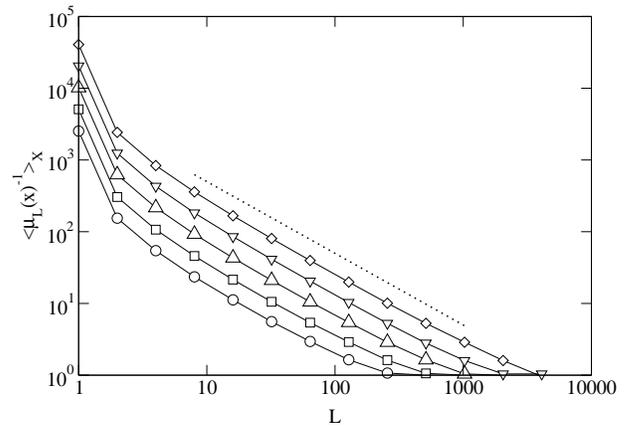}}
\caption{Average $\langle\ln(\mu_L^{-1})\rangle$. the dashed line
corresponds to a function $\sim L^{-1}$. Symbols and averaging as in
Fig.~\protect\ref{fig:d2}.}
\label{fig:d0}
\end{figure}

\section{Dimension analysis of the Klemm-Egu{\'\i}luz (KE) network.}
We have performed extensive numerical studies on the KE structured
scale-free networks \cite{Klemm02a}. The procedure to generate
them is as follows. Start with $m$ active, fully-connected nodes.
At each time step add a new node and attach $m$ new links between
the new node and the old active nodes. Activate the new node and
deactivate one of the active nodes according to the
probability $\Pi(k) \propto k^{-1}$, where $k$ is the degree of
the node. This algorithm
generates networks with a power law degree distribution $P(k) \sim
k^{-\gamma}$, where $\gamma \simeq 3- 1/m$ \cite{KlemmThesis}.
In Fig.~\ref{fig:d2}
we plot the average normalized number of nodes at distance $L$,
$\langle \mu_L (X_i) \rangle$ for different system sizes, with $m=3$.
By using
(\ref{moments}) we find that the best fit gives the correlation
dimension $D_2 = 0.92$. In Fig.~\ref{fig:d1} we plot the average
slope $\langle \ln \mu_L (X_i) / \ln L\rangle$ for different
system sizes. From (\ref{D1}) the plateau regime indicates an
average value of the information dimension $D_1 = 0.97$. Finally
from Fig.~\ref{fig:d0}, the best fit gives, via (\ref{moments}),
the capacity dimension $D_0 = 1.0$.

The differences in the values of $D_q$ for different $q$  are
a consequence of the inhomogeneity of the network. We will not
concentrate here on this interesting point, but focus on the fact
that the dimension estimates are finite and
close to 1. This confirms the conclusions in \cite{Klemm02b,Vazquez02},
obtained from the scaling of the network's diameter, and opens the
way for future characterization of the whole spectrum of
dimensions in this and other network models. We can say that the
network behaves, in the sense of the definitions introduced above,
very close to one-dimensional. It should be stressed that {\em a priori}
there has not been an obvious Euclidian space containing the
structure from which to calculate the dimensions $d_q$. In the following we
consider dynamical processes occurring in KE networks and
interpret them to the light of the dimension study.

\begin{figure}
\centerline{\epsfig{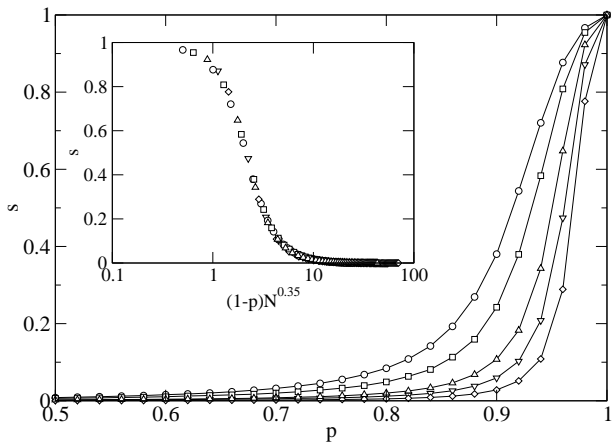}}
\caption{Average relative size of the largest cluster $s$ in site
percolation for different occupation probabilities $p$ in $m=3$ KE
networks. The inset shows a finite size scaling
$s=F((1-p)N^{0.35})$. The average values have been obtained from
1000 percolation realizations in 10 different networks. Network
size $N=$10000 (circles), 20000 (squares), 50000 (triangle up),
100000 (triangle down), 200000 (diamonds). }
\label{fig:per}
\end{figure}

\section{Site percolation.}
An important process occurring on real networks is the propagation
of information, or of diseases. If the nodes of the network are
either susceptible to the disease with probability $p$ or immune
to it with probability $1-p$, and the disease propagates through
the links, the maximum number of individuals that can be affected
by an epidemic outbreak is given by the size of the largest
connected cluster in site percolation, where occupation of a site
means susceptibility to the disease. We have performed numerical
simulations of such a process in KE networks. In Fig.~\ref{fig:per}
we show the average relative size $s$ of the connected cluster
around an occupied site for different system sizes and $m=3$. The
results indicate that for any system size $N$ there is a (broad)
transition to the percolating state at some value of the
occupation probability $p_c$. However it is seen that $p_c \to 1$
as $N \to \infty$. Thus, the percolating transition occurs at
$p_c=1$ in the infinite-size limit. This is precisely the expected
behavior in a one-dimensional structure. The inset shows that the
relative size of the largest cluster scales as
$F((1-p)N^{\alpha})$, where the exponent $\alpha$ depends on the
average connectivity of the network. This finite size scaling
behavior is also what one finds for percolation in one-dimensional
regular lattices. In addition we have found that $\alpha=0.35$ for
$m=3$, while $\alpha\simeq0.21$ for $m=5$, suggesting that
$\alpha$ is a decreasing function of the average degree $\langle
k\rangle = 2m$. These numbers are in good agreement with site
percolation in one-dimensional lattices with radius of interaction
$z$ (i.e. two occupied sites are considered connected if they are
$z$ or less sites apart, so that the degree of each site is $2z$).
There one can show that $\alpha = z^{-1}$.

It is worth noting
that this result is in contrast with the zero percolation
threshold found \cite{Pastor01} in random scale-free networks of
the Barab{\'a}si-Albert type \cite{Barabasi99}. For such networks,
we have checked (for $q=0,1,2$) that the moments $\langle
\mu^q\rangle$ do not scale as a power law of $L$, but instead $D_q
\rightarrow \infty$ in this case. Thus, the different behavior may
be associated to the different dimensionality.

\section{Conclusions.} We have introduced definitions of dimensions useful
for the study of complex networks. In particular, we have calculated
the capacity, information and correlation dimension of a type of
hierarchically structured scale-free network. We have shown that some
dynamical properties of this class of networks can be understood in the
light of the dimension analysis. It would be worth to search for other
complex networks displaying finite dimension spectra. We think that
possible candidates would be those with some underlying ``regular''
topology \cite{Rozenfeld02,Warren02}.

We acknowledge financial support from MCyT (Spain) and FEDER (EU)
through Projects CONOCE BFM2000-1108 and BFM2002-04474-C02-01.

\end{document}